\documentclass[journal]{IEEEtran}

\usepackage[pdftex]{graphicx}
\usepackage{graphicx}
\DeclareGraphicsExtensions{.pdf,.jpeg,.png}
\usepackage{subfigure}
\usepackage[cmex10]{amsmath}
\usepackage{algorithmic}
\usepackage{algorithm}
\usepackage{array}
\usepackage{mdwmath}
\usepackage{mdwtab}
\usepackage{eqparbox}
\usepackage{url} 
\usepackage{import}
\usepackage{tabularx}
\usepackage{enumerate}
\usepackage{booktabs}
\usepackage{multirow}
\usepackage[separate-uncertainty,tight-spacing=true]{siunitx}
\usepackage{textcomp}
\usepackage{xcolor}
\usepackage{hyperref}
\usepackage{notoccite}
\usepackage{cleveref}

\begin{document}
\bstctlcite{IEEEexample:BSTcontrol}
    \title{PlexiChain: A Secure Blockchain-based Flexibility Aggregator Framework}
  \author{Samuel~Karumba,~\IEEEmembership{Student Member,~IEEE;}
      Salil~S.~Kanhere, ~\IEEEmembership{Senior Member,~IEEE;}
      Raja~Jurdak, ~\IEEEmembership{Senior Member,~IEEE;} 
      and~Subbu~Sethuvenkatraman

 \thanks{Samuel Karumba and Salil Kanhere are with the school of Computer Science and Engineering, UNSW, Asutralia. E-mail:{\{s.karumba, salil.kanhere\}@unsw.edu.au}}
  \thanks{Raja Jurdak is with the school of Computer Science, QUT, Australia E-mail: r.jurdak@qut.edu.au}
  \thanks{Subbu Sethuvenkatraman is with Energy Business Unit, CSIRO, Australia E-mail: subbu.sethuvenkatraman@csiro.au}
  
}

\markboth{IEEE TRANSACTIONS ON SMART GRID
}{Samuel \MakeLowercase{\textit{et al.}}: Energy Efficiency}

\maketitle

\begin{abstract}
Flexible resources in built environments are seen as a low-cost opportunity for delivering grid management services. 
Consequently, the centralised aggregator model, where the aggregator is used to bundle demand flexibility from flexible resources and deliver it to flexibility customers such as Distributed/Transmission System Operator (DSO/TSO) in flexibility markets, has been adopted. 
However, the aggregator role introduces various security and trust challenges. 
In this work, we propose a blockchain-based flexibility trading framework dubbed PlexiChain to address the security and trust challenges the aggregator poses in the centralised aggregator model.
The security evaluations performed using a real-world dataset show that PlexiChain is robust against known security attacks, such as MadIoT and False Data Injection attacks.
Additionally, the performance evaluations show that PlexiChain has lower computation and communication costs than other blockchain-based applications in resource-constrained environments. 
\end{abstract}


%
\IEEEpeerreviewmaketitle



\section{Introduction} \label{sec:introduction}
\subsection{Overview}
Buildings and other structural facilities in built environments account for more than 40\% of the world's total energy use and are responsible for nearly 15\% of the world's greenhouse gas emissions \cite{IEA2022}.
Meanwhile, Demand Flexibility (DF) can help reduce energy consumption and greenhouse gas emissions in built environments \cite{Qi2017a}.  
DF is defined as the ability of end-users' to "shape, shift, shed, and shimmy" their flexible loads to deliver grid management services under various time scales, often for financial incentives \cite{DOE2021}.
Flexible loads in built environments include: 
(i) HVAC (Heating, Ventilation, Air-conditioning, and cooling) system. 
(ii) Hot Water Systems (HWS). 
(iii) Energy Storage Systems (ESS) (e.g. batteries, electric vehicles, and thermal storage). 
\begin{figure}
    \centering
    \includegraphics[width=\linewidth]{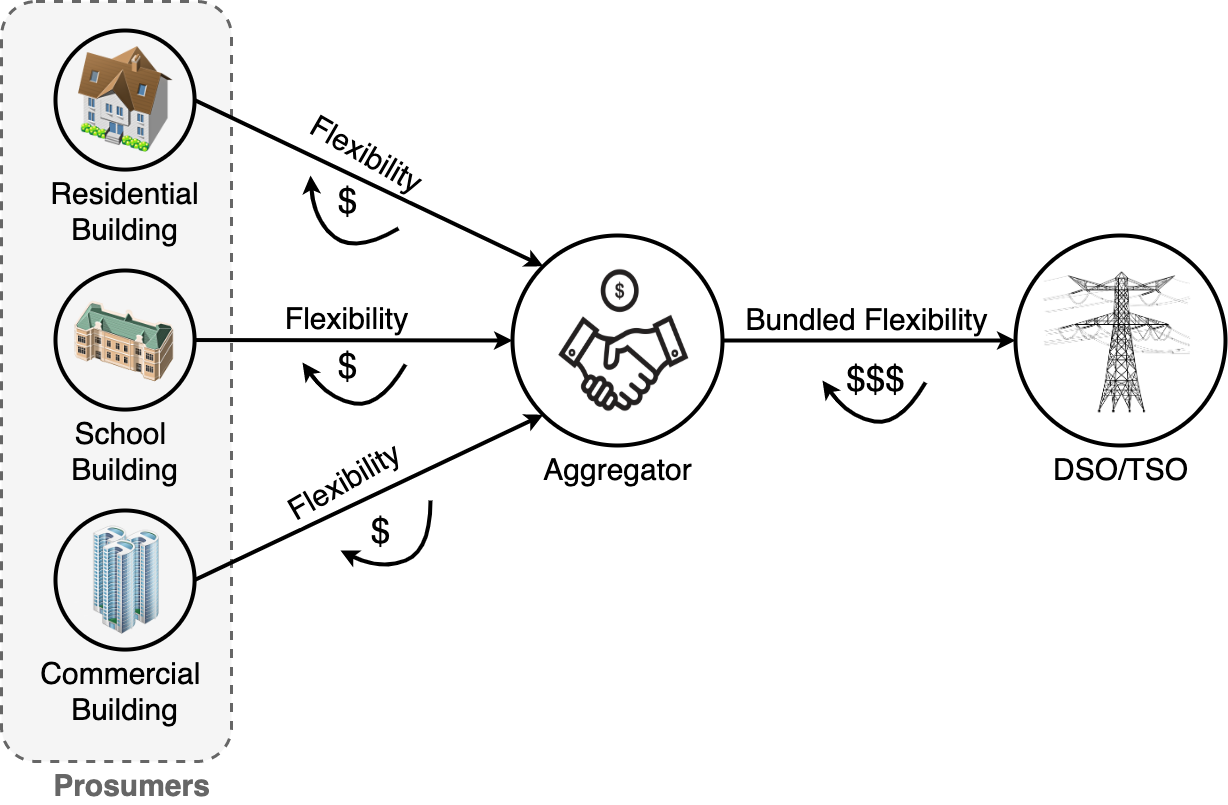}
    \caption{Centralised Aggregator Model}
    \label{fig:incentive-based_dr_program}
\end{figure}

Moreover, with the increasing integration of Distributed Energy Resources (DERs), end-users are transitioning from energy consumers to energy prosumers \cite{Karumba2020}.
Prosumers are end-users who not only consume energy but can also produce energy. 
Therefore, prosumers can provide DF by varying their on-site DERs, such as rooftop solar panels, wind turbines, battery storage, and electric vehicles (EV).
Although prosumers in built environments have the potential to provide this new form of DF, the flexibility offered individually may be small.
Consequently, the centralised aggregator model, where the aggregator bundles demand flexibility from prosumers and offer it to the flexibility customers, is proposed in \cite{USEFFoundation2016}, as illustrated in \Cref{fig:incentive-based_dr_program}.
Often, the aggregator is a Trusted Third Party (TTP) firm, while the flexibility customers are transmission system operators (TSOs), distribution system operators (DSOs) and balance responsible parties (BRPs) \cite{USEFFoundation2016}.

Although the centralised aggregator model enables individual prosumers to participate in the flexibility markets, using a TTP as the aggregator introduces a list of security and trust challenges \cite{USEFFoundation2016}.
These challenges include: 
\begin{enumerate}
    \item Ensuring data integrity during collection, transmission, and storage from individual prosumers' devices to the aggregator.
    \item Guaranteeing trustworthiness in remote access and control of flexible resources while considering the prosumers and other building occupants' preferences and behavioural patterns.
    \item Finding a balance between transparency and confidentiality in the information exchanged during flexibility trading.
\end{enumerate}
In this paper, we seek to address the security and trust challenges mentioned above using blockchain technology.

\subsection{State-of-the-art}
Recently, blockchain technology has emerged as a new technology that promises to provide the salient features of decentralisation, trust, security, and transparency for applications in the energy sector \cite{Karumba2020}.
The blockchain's salient features are provided by its suite of technologies, which include distributed ledger, consensus algorithms, Public Key Cryptography (PKC), and smart contracts, respectively \cite{Karumba2020}.
A smart contract is a deterministic piece of code used to automate the execution of a contract without the need for a TTP \cite{Dorri2019a}.

Although blockchain has numerous applications in the energy sector, its suite of technologies incurs high computation and communication costs, hindering its application in resource-constrained environments such as Internet-of-Things (IoT) networks \cite{Dorri2021a}.
Smart grids are inherently IoT networks with numerous resources constrained computing devices, which limit blockchain application in the energy sector \cite{Karumba2020}.
Consequently, researchers in the IoT domain have proposed the use of: removable ledgers \cite{Dorri2021a}, scalable networks \cite{Karumba2020}, and lightweight consensus mechanisms \cite{Dorri2019a} for effective blockchain application in IoT networks. 

While the outlined research works have significantly reduced the computation and communication costs in blockchain, the efficiency of its PKC has not been fully explored.
Indeed, for PKC to be valuable, all blockchain participants must have registered identities stored in a standard X.509 digital certificate format \cite{Guo2020a}. 
Digital certificates rely on the costly Public Key Infrastructure (PKI), which ensures that the digital certificates are valid and trustworthy.
However, there are multiple hurdles to relying on the certificate-based PKI, including: (i) PKI requires Certificate Authorities (CAs), which act as a TTP for issuing and verifying digital certificates. 
(ii) PKI requires a complex certificate management process, which incurs high computation and communication costs in resource-constrained IoT networks \cite{Dorri2019a}. 
Therefore, we must increase the efficiency of the PKC scheme for effective blockchain applications in IoT.

\subsection{Our Contributions}
In this work, we present a decentralised aggregator model, where the blockchain's suite of technologies is used to address the security and trust challenges associated with the centralised aggregator model.
In our model, we propose using PlexiChain, a Blockchain-based Flexibility Aggregator Framework, to act as the aggregator.
PlexiChain adopts the Constraint Satisfaction Problem (CSP) concept to model the aggregator role without needing a TTP.
A CSP is a mathematical model that can be used to solve problems specified by a set of constraints \cite{Brailsford1999}.

Additionally, PlexiChain extends our previous work on the Hypergraph-based Adaptive consoRtium Blockchain (HARB) proposed in \cite{Karumba2020} to include a certificate-less PKC (CL-PKC) for the application of blockchain in smart grids. 
HARB is a lightweight, scalable blockchain framework for application in IoT networks.
In the proposed CL-PKC, we argue that storing registered user identities in Non-Fungible Tokens (NFTs) instead of the standard X.509 digital certificates could significantly improve the efficiency of blockchain for applications in IoT networks.
The specific contributions of our work are as follows:
\begin{enumerate}
    \item We propose an NFT-based PKI to ensure the integrity of data available from IoT devices during collection, transmission and storage.
    \item We propose a CSP-based demand Flexibility Aggregator Smart Contract (dFASC) to automate remote access and control of flexible resources while considering building occupants' preferences and behavioural patterns.
    \item We propose a privacy-preserving Decentralised Application (DApp) to find a balance between transparency and confidentiality in DF Trading.
    \item Lastly, we present quantitative security and performance evaluations of the proposed PlexiChain platform.  
\end{enumerate}

The rest of the paper is organised as follows:
In \Cref{sec:related_work}, we review the related work.
\Cref{sec:plexichain} presents a detailed description of the proposed PlexiChain system.
In \Cref{sec:security_evaluation}, we discuss the security and performance evaluations of the proposed PlexiChain framework.
Finally, \Cref{sec:conclusions} draws our concluding remarks.

\section{Related Work} \label{sec:related_work}
This section reviews related works on blockchain-based decentralised aggregator models and certificate-less PKC schemes.
Additionally, this section describes NFTs and CSP in more detail to provide a preliminary of the proposed platform's building blocks.

\subsection{Blockchain-based Decentralised Aggregator models}
In the proposed blockchain-based decentralised aggregator models, blockchain plays the aggregator role.
Moreover, the use of blockchain goes beyond the decentralisation of the aggregator model in terms of providing transactions security, trust, transparency, and provenance, as thoroughly reviewed in \cite{Khajeh2020a}. 
For instance, Pop et al. \cite{Pop2018a} investigated the use of blockchain technology for secure, reliable and timely delivery of energy flexibility to all market participants in a Local Flexibility Market (LFM).
The authors also used smart contracts to validate DF agreements and transparently balance demand and production in their approach.
However, blockchains trade-off transaction confidentiality in exchange for transparency \cite{Karumba2020}. 

In \cite{Advanced2018}, the authors developed a blockchain-based decentralised aggregator platform that ensures sensitive data remains safe and secure.
Similarly, Van Cutsem et al. \cite{VanCutsem2020a} used smart contracts to guarantee openness in monitoring the expected flexibility level among untrustworthy participants (i.e. prosumers, DSO, TSO, and BRPs) involved in the flexibility market.
Although blockchain ensures data security, trust, and transparency in these blockchain-based platforms, blockchain does not provide component-level security, especially in resource-constrained environments.
For instance, only high computing devices (e.g. a personal computer) are selected to act as peers on the blockchain network since the blockchain's suite of technologies incur high computation and communication costs.
Consequently, the integrity of IoT data collected, transmitted, and stored in the blockchain system cannot be guaranteed.

Therefore, our work aims at extending the HARB framework to include an NFT-based PKI to guarantee the integrity of IoT data sources in resource-constrained environments such as smart grids.
Additionally, we model the resource-constrained environment as CSP to ensure safety in controlling the flexible resources via the IoT devices.
Moreover, we provide a privacy-preserving mechanism to provide a balance between transaction confidentiality and transparency.

\subsection{Certificate-less PKC}
Although various researchers have proposed lightweight security mechanisms for the application of blockchain in IoT as summarised in \cite{Ali2019a}, none of them has applied the CL-PKC directly to the blockchain's PKI. 
Consequently, Guo et al. \cite{Guo2020a} attempted to introduce CL-PKI into the consortium blockchain.
The authors proposed a key-derived Controllable Lightweight Secure Certificate-less Signature (CLS$^{2}$) algorithm in their work.
A key feature of their CL-PKC system is that it does not require a digital certificate or a public key.
Nonetheless, the proposed CLS$^{2}$ scheme does not meet the required security robustness to prevent confidentiality, integrity and availability attacks.
This is because if the private key is compromised, an attacker can assume the identity of an authorised participant.

In this work, we have proposed an NFT-based PKI to provide unique digital identities for resource-constrained devices. 
An NFT is issued and maintained on the blockchain network. 
Therefore, there is no need for a TTP to verify their authenticity.

\subsection{None-Fungible Tokens (NFT)} \label{sec:nft}
An NFT is a type of unique cryptocurrency which cannot be exchanged like-for-like, making it suitable for identifying something or someone in a unique way \cite{Wang2021}.
An IoT device $v_i$ is considered uniquely identifiable if it's bound to an NFT $E_i$.
The token structure is described as a tuple to provide a compact representation:
\begin{equation}
    \begin{aligned}
        E_i=<Token ID, Token Name, Device ID,\\ Pseudo ID, Owner ID, Constraints,\\ Issue Time>
    \end{aligned}
\end{equation}
Where \textit{DeviceID} is the ID of a physical IoT device generated by the Physically Unclonable Function (PUF).
We assume that all IoT devices are capable of implementing PUF.
The token structure is blockchain agnostic and can be defined using any standard, including the ERC-721 standard in Ethereum and the Unspent Transaction Output (UTXO) standard in Bitcoin and Hyperledger Fabric.

\subsection{Constraint Satisfaction Problems (CSP)}\label{sec:csp}
A CSP $P$ is formally described by a triplet $P$:
\begin{equation}
    \begin{aligned}
        P=(V,D,C)
    \end{aligned}
\end{equation}
Where $ V=\left \{ v_{1},v_{2}, \cdots v_{n}\right \}$  is a finite set of variables with $n=|V|$ being the number of variables to be solved in the problem, $ D=\left \{ d_{1},d_{2}, \cdots d_{n}\right \} $  is a finite set of domains for variables in $V$, and $C=\left \{ c_{1},c_{2}, \cdots c_{m}\right \}$ is a set of constraints.
There are various types of constraints in CSP, including unary, binary, and high-order.
Graph-theoretical notation can describe the structural properties of CSP models in which nodes represent variables and arcs represent constraints.
Most problems are represented as binary CSP, where the constraints involve pairs \cite{Zabih1990}.
A binary CSP has an associated constraint graph $P=(V,E)$, where $V$ is a set of variables, and $E=\{ \left ( s_{i} ,\rho  \right )| s=\left ( v_{i},v_{j} \right )$ is a set of binary constraint relations between $ v_{i}$ and $v_{j}$ for all $v_{i}\wedge v_{j} \in V\}$.
However, the binary constraint graph hides the contextual interaction information, particularly the tightness of the constraints.
Consequently, the standard notation of hypergraphs can be adopted to represent high-order relationship constraints \cite{Bretto2013a}.
For any CSP instance $P$, its associated constraint hypergraph is a graph $H(P)$:
\begin{equation}
    \begin{aligned}
        H(P)  =  \left ( V,E,C \right )
    \end{aligned}
\end{equation}
Where $V$ is a set of variables, $E$ is a set of domains, and $C=\{ \left ( s_{i} ,\rho  \right )| s_{i}=(v_{1},v_{2},v_{3},\dots,v_{t})\}$ is a set of constraints.
Each constraint $c \in C$ is a high-order relation  on $s_{i}$ given the assignment function $\rho(s_{i})$. 

Next, in \cref{sec:plexichain}, we present the proposed PlexiChain platform, which implements the security requirement based on blockchain and CSP to address the security and trust challenges mentioned in \Cref{sec:introduction}.
\section{PlexiChain} \label{sec:plexichain}
This section presents a detailed overview of the proposed PlexiChain system.
First, we present the PlexiChain architecture, which extended the HARB framework.
The architecture gives a high-level overview of the underlying network layers and system components.
Then, in the following subsections, we describe the proposed lightweight NFT-based PKI, dFASC, and DF Trading DApp. 

\begin{figure}[!th]
    \centering
    \includegraphics[width=\linewidth]{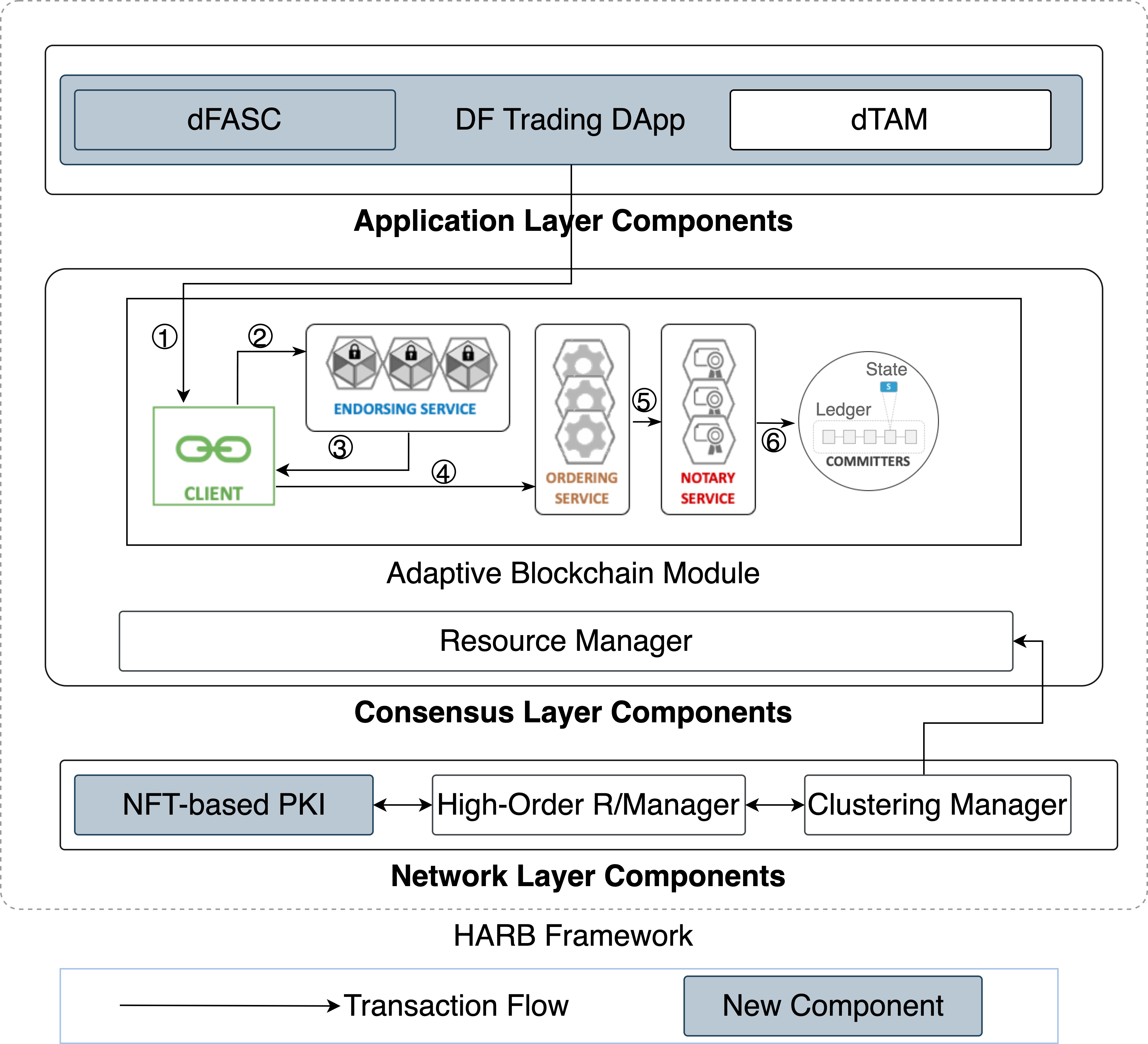}
    \caption{PlexiChain System Architecture}
    \label{fig:plexichain}
\end{figure}

\subsection{System Architecture}
The HARB framework is a custom Hyperledger Fabric blockchain framework that exploits the rich representation ability of hypergraphs to scale the existing blockchain framework \cite{Karumba2020}.
The architecture is split into three abstraction layers, namely the underlay (network), the Overlay (consensus), and the control (application) layers, as illustrated in \Cref{fig:plexichain}.
Each layer outlines a set of lightweight blockchain components for applications in resource-constrained IoT environments.
To design the PlexiChain architecture, we extend each layer as follows:

\subsubsection{Network layer}
The network layer comprises a high-order relationship manager and a network clustering manager.
The High-Order R/Ship manager maps out complex relationships between physical nodes in the network.
The clustering manager then uses these complex relationships to organise nodes into maximally coherent groups or clusters.
Interactions within the clusters are saved in a local data store for data confidentiality, while inter-cluster interactions are recorded on the blockchain ledger for transparency.
In addition, we extend the network layer to include the proposed Lightweight NFT-based PKI for ensuring the integrity of IoT data during collection, transmission and storage.
A detailed description of the proposed Lightweight NFT-based PKI is given in \Cref{sec:light_id}.

\subsubsection{Consensus layer}
The consensus layer hosts a set of computing devices grouped into components used to execute the various blockchain services, such as endorsing, ordering, notarising, and committing.
The Adaptive Blockchain Module (ABM) is used to dynamically assign blockchain service roles to the computing devices based on their computing resource levels.
Collectively, blockchain services are used to guarantee trustworthiness in a decentralised system.
All transactions are first validated and endorsed by the endorsing peers since smart contracts are deployed on the endorsing peers. 
Second, the ordering peers verify and order all transaction endorsements through a consensus mechanism.
However, for inter-cluster interactions, transaction endorsements are verified by the notary peers and ordered by the ordering peer.
Lastly, all transactions are immutably stored by the committing peers in a distributed ledger.

\subsubsection{Application layer}
In \cite{Karumba2020}, the application layer provides the end-user with the system's interaction interfaces such as dTAM, SDKs, and APIs. 
dTAM is an application agnostic smart contract for ensuring data privacy for cross-cluster transactions.
We extend the application layer in \cite{Karumba2020} to include a decentralised application (DApp) for DF trading (DF-DApp) and a decentralised Flexibility Aggregator Smart Contract (dFASC). 
A detailed description of the proposed DF-DApp is presented in \Cref{sec:df_trading_dapp} while that of dFASC is presented in \Cref{sec:df_agg_sc}.

\subsection{Lightweight NFT-base PKI} \label{sec:light_id}
In the PlexiChain architecture, a Lightweight NFT-base PKI is proposed based on the unique NFTs introduced in \Cref{sec:nft}. 
Unlike the digital certificates issued and verified by a TTP, the NFTs are issued by a deterministic smart contract and are verified by multiple nodes on the distributed blockchain network.
The NFT-PKI protocol for key generation, issuance of digital identity, and digital signature creation and verification have been described in two phases: the Enrollment and the Authentication Phases.

\subsubsection{Enrollment Phase}
The enrollment phase describes the process of generating an NFT as a trusted digital identity for an IoT device $v_i$.
The various algorithms in the enrollment phase are described as follows:

\begin{enumerate}
    \item \textbf{Setup:} 
    An IoT device $v_i$ initiates the process by sending an \textbf{Enrolment Request} to the anchor peer.
    The anchor peer then initialises the PKE.Setup() algorithm takes a security parameter $\lambda$ as input and outputs its public-private $(msk,mpk)$ key pair.
    The anchor peer then associates a hash function $H$ with its private key $msk$ to generate a challenge $C$, sent to the PUF inside the IoT device.
    \item \textbf{PUF():}
    Like humans, each IoT device has its unique fingerprint created during manufacture.
    This intrinsic IoT device characteristic can be extracted by adding a specific circuit called the PUF circuit \cite{Babaei2019a}. 
    A PUF is a physical object that, for a given challenge (C), provides a response (R) that serves as a unique identifier for a given IoT device \cite{Arcenegui2021}.
    
    
    \item \textbf{NFT generation:}
    The NFT generation algorithm describes generating the public-private key pair and binding the device's public key to an NFT token.

    \begin{algorithm}[!h] 
    \caption{Identity generation and Binding procedure}
    \label{alg:key_generation}
        \begin{algorithmic}[1]
            \renewcommand{\algorithmicrequire}{\textbf{Input:}}
            \renewcommand{\algorithmicensure}{\textbf{Output:}}
            \REQUIRE $R,msk$  
            \ENSURE $sk_i,TokenID_i$
            \STATE $1,\perp \leftarrow Invoke(query,R)$
            \IF{($flag==1)$}
            \STATE $sk_i,pk_i\leftarrow PKE.Gen(msk)$
            \STATE $NFT\leftarrow Invoke(createNFT,R,ownerID)$
            \STATE $NFT_i \leftarrow BindIoTtoNFT(pk_i)$
            \ELSE
            \RETURN $\perp$
            \ENDIF
            \RETURN $sk_i,NFT_i.TokenID$ 
        \end{algorithmic} 
    \end{algorithm}

    First, the algorithm checks if the device $v_i$ has been enrolled before by Invoking the NFT SC with function name $query$ and unique identity $R$ as the input parameters.
    If the device has not been enrolled before, the NFT SC returns a $flag=1$, else it returns a $flag=\perp$, where $\perp$ is an error message.
    Second, the algorithm runs the PKE.Gen() function, which associates the anchor peer's private key $msk$ with a hash function $H()$ to generate the devices public-private key pair $pk_i,sk_i$ for device $v_i$.
    Third, the algorithm invokes the NFT SC and passes the function name $createNFT$, unique identities $R, OwnerID,pk_i$ and other device metadata necessary to create an NFT $E_i$, as inputs.
    Like digital signing certificates, NFTs can be used to identify devices and implement access control mechanisms for sensitive or valuable information, as described in \Cref{sec:nft}. 
    Lastly, the algorithm binds the NFT to the IoT device by associating its public key $pk_i$ with the generated NFT $E_i$.
    The NFT $E_i$ is stored on the blockchain network, while the public key and the token ID $sk_i, TokenID_i$ are sent to the IoT device through a secure channel.
    The IoT device securely stores its private key in its registry.
\end{enumerate}

\subsubsection{Authentication phase}
In the authentication phase, the IoT device proves its identity to a verifier (transacting peer) on the blockchain network. 
The following set of algorithms describes the authentication protocol:

\begin{enumerate}
    \item \textbf{Sign():} 
    The transacting device $v_i$ generate a digital signature $\sigma$ using its private key $sk_{i}$ and a message $m\in \{0,1\}$ as inputs.
    The signing process is described in \Cref{alg:sign}.
    \begin{algorithm}[!h] 
    \caption{Message signing procedure}
    \label{alg:sign}
        \begin{algorithmic}[1]
            \renewcommand{\algorithmicrequire}{\textbf{Input:}}
            \renewcommand{\algorithmicensure}{\textbf{Output:}}
            \REQUIRE $m=\{1,0\}^*,sk_i$  
            \ENSURE $\sigma$
            \STATE $\sigma \leftarrow PKE.Sign(m,sk_{i})$
            \RETURN $\sigma$ 
        \end{algorithmic} 
    \end{algorithm}
    The IoT device then dispatches the digital signature $\sigma$ and its $TokenID$ to the recipient for verification and processing on the blockchain network.
    \item \textbf{Verify:} 
    The verifier executes this algorithm described in \Cref{alg:verify}.
    The algorithm takes the digital signature $\sigma$ and the token ID $TokenID$ as input and returns $1$ if verified or $0$ if the verification fails.
    \begin{algorithm}[!h] 
    \caption{Message verification procedure}
    \label{alg:verify}
        \begin{algorithmic}[1]
            \renewcommand{\algorithmicrequire}{\textbf{Input:}}
            \renewcommand{\algorithmicensure}{\textbf{Output:}}
            \REQUIRE $\sigma,TokenID_i$  
            \ENSURE $1/0$
            \STATE $NFT,\perp \leftarrow Invoke(query,R)$
            \IF{($TokenID == NFT.TokenID)$}
            \FORALL{$i \in \{0,1\}$}
            \STATE $C\leftarrow H'(i,msk)$
            \ENDFOR
            \ELSE
            \RETURN $\perp$
            \ENDIF
            \STATE $R \leftarrow PUF(C)$
            \IF{($R == NFT.R)$}
            \STATE $1/0 \leftarrow PKE.Verify(m,NFT.pk_i)$
            \ELSE
            \RETURN $\perp$
            \ENDIF
            \RETURN $1/0$ 
        \end{algorithmic} 
    \end{algorithm}
    The recipient first retrieves the IoT device's NFT token for identity verification by invoking the NFT SC with a query request.
    The query SC function takes the $TokenID$ as input.
    If the NFT with token Id $TokenID$ does not exist, the algorithm halts and returns an error message.
    If the NFT exists, the SC returns the NFT to the verifier. 
    The verifier then parses the NFT token to retrieve the unique identity $R$ of the IoT device.
    To verify that the message was sent by the device  $v_i$, the algorithm sends a challenge $C$ to the PUF in the IoT device.
    The PUF uses the challenge to generate the unique response $R$.
    If the response $R$ matches the one recorded on the NFT token, the device identity has been verified.
    Lastly, the algorithm uses the device public key $NFT.pk_i$ to verify the message signature. 
    Once the signature has been verified, the verifier can process the message. Else if not valid, it returns an error message $\perp$.
\end{enumerate}
Since all the devices in the network can authenticate and verify each other on the network, they can autonomously operate without the need for TTP or external intervention. 
The next section introduces the proposed blockchain-based aggregator model for automating remote access and controlling flexible resources without TTP.

\subsection{Privacy-preserving DF Trading DApp} \label{sec:df_trading_dapp}
The DF-DApp provides an interface for market participants to utilise the various blockchain services such as security, trust, decentralisation, and transparency in DF trading.
The DF-DApp is divided into various components, including the data model, smart contracts, and workflows.
Each component is described below.

\subsubsection{Data Model}
A data model is a standardised structure that describes how the real-world entities and elements of data relate to one another. 
In our data model, the main entities and model include the Flexibility Requests, Workflows, Notifications and Actors.
Each element/entity is described as follows:
\begin{itemize}
    \item \textbf{Flexibility Requests:} This model maintains a record of all demand flexibility service requests issued by the DSO/TSO for reliable grid management.
    The requests contain information about the flexibility service required, such as unique identity, flexibility type (energy demand or supply), flexibility amount, and the incentive. 
    \item \textbf{Workflows:} A workflow model maintains a list of workflows defined to govern a certain business process. 
    Each workflow contains information such as unique identity, actors (participants), events, and state.
    \item \textbf{Notifications:} A notification is triggered by a workflow event. 
    The relevant participants in a workflow register listeners to receive generated notifications.
    \item \textbf{Actors:} An actor is a participant with a specific role in the workflow. 
    Actors initiate the workflow activities and receive event notifications.
    Smart contracts can also be actors in the system performing predefined tasks in the workflow.
\end{itemize}

\subsubsection{Smart contracts}
In the Hyperledger Fabric, which is the underlying blockchain platform for the PlexiChain framework, the implementation of a smart contract is referred to as a \textit{chaincode} \cite{Androulaki2018}.
Two main chaincodes are utilised in PlexiChain, namely dFASC and dTAM.
The dTAM chaincode proposed in \cite{Karumba2020}, is responsible for workflow and data privacy management, while dFASC is responsible for the decentralisation of the centralised aggregator model, as described in \Cref{sec:df_agg_sc}.

\subsubsection{Workflow}
A workflow is a step-by-step description of the activity sequence (business rules) that the market participants must perform to fulfil a business transaction. 
The activities are modelled as internal events triggered by the actors. 
A workflow event is associated with a \textbf{Notification} to advance the state of a workflow.
Steps in the flexibility trading workflow are outlined as follows:

\begin{enumerate}
    \item Create Flexibility Request. 
    The flexibility customer (e.g. DSO/TSO) initiates this  activity by invoking the \textit{createFlexRequest()} function on dFASC.
    The invoked function triggers a \textit{CREATE\_FLEX\_REQUEST} event, which is sent to the dTAM smart contract. 
    The dTAM smart contract updates the workflow event state in the blockchain ledger and publishes a \textit{$flex\_bid\_request$} notification.
    All prosumers subscribe to receive the published \textit{$flex\_bid\_request$} notifications.
    \item Place Flexibility Bids. 
    On receipt, the prosumers evaluate the flexibility bid request notification and place flexibility bids by invoking the \textit{flexBidding()} function on dFASC with the amount of flexibility available to offer.
    The invoked function triggers the \textit{BID\_OFFER} event, which is sent to the dTAM smart contract. 
    dTAM updates the workflow events state in the ledger and publishes a \textit{bid\_offer} notification. 
    dFASC subscribes to receive the published \textit{bid\_offer} notifications.
    \item The dFASC chaincode receives the \textit{bid\_offer} notifications and aggregates the flexibility offered by multiple prosumers in their bids.
    If the objective function presented by the objective function in \Cref{eq:objective_func} is solved, the market is cleared, and the bidding process is closed.
    The dFASC then invokes the \textit{scheduleDR()} function on dTAM, which triggers the CREATE\_DF\_SCHEDULING event sent to the dTAM smart contract.
    dTAM updates the workflow state in the blockchain ledger and publishes a \textit{$df\_scheduling$} notification.
    A \textit{$df\_scheduling$} notification describes a set of constraints to generate or curtail electric energy (depending on the constraint set points) for a specified period $\Delta_t$.
    All flexible resources subscribe to the \textit{$df\_scheduling$} notification.
    \item When the specified time $\Delta_t$ elapses, dFASC invokes the \textit{activationAndSettlement()} function.
    The invoked function triggers the ACTIVATION\_SETTLEMENT event, which is sent to the dTAM.
    dTAM updates the workflow state in the blockchain ledger and publishes a \textit{$df\_fulfilled$} notification.
    All flexibility customers subscribe to the \textit{$df\_fulfilled$} notification
\end{enumerate}

A different workflow handles the flexibility requests incentive settlements.
Currently, the incentive settlements workflow is outside the scope of this work, and it is assumed that the settlements are done outside the PlexiChain system.

\subsection{DF Aggregator Smart Contract (dFASC)} \label{sec:df_agg_sc}
Consider the centralised aggregator model illustrated in \Cref{fig:incentive-based_dr_program}.
Let $P$ denote the set of prosumers.
We assume that each prosumer $p \in P$ is equipped with a smart meter, which has a local controller \cite{Chuang2008}.
A smart meter is an IoT device that can adjust the consumption or generation set points of flexible resources.
All smart meters are connected to the aggregator for remote control and monitoring.
To automate remote control and monitoring of flexible resources while considering the prosumers and other building occupants' preferences and behavioural patterns, we implement the high-order CSP model presented in \Cref{eq:csp_model} as dFASC. 
\begin{equation} \label{eq:csp_model}
    H=(V,E,C)
\end{equation}
$H$ is the objective function, $V$ is a set of flexible resources, $E$ is a set of flexible resources domain models, and $C$ is a set of DSO request constraints. 
The formulation of each variable can be described as follows:

\subsubsection{Objective function $H$}
Reflects the minimisation of dFASC's operational cost of meeting a DF service request from a flexibility customer (i.e. DSO or TSO).
The objective function is given by:
\begin{equation}
    H = min\int_{DF_{t}^{s,q,d}}^{\$}
    \label{eq:objective_func}
\end{equation}
Where each DF request \textbf{($DF_{t}^{s,q,d}$)} and its incentive $(\$)$ comes from \textbf{dFASC-to-DF\_Customer} contract and is predefined before the execution phase. 
The set parameters $t,s,q$ and $d$ represent the requested DF service's time, shape, quantity, and direction constraints.

\subsubsection{Flexible resources set $V$}
Let $V_{p}=\{v_{1},v_{2},v_{3},\dots,v_{n}\}$ denote a set of flexible resources owned by a prosumer $p$.
Therefore, the variable set $V=\left \{ V_{p_{1}},V_{p_{2}},V_{p_{3}}, \cdots ,V_{p_{x}} \right \}$ contains all flexible resources owned by $x$ number of prosumers.
These flexible resources are remotely controlled and monitored by the dFASC to meet the objective function $H$.  

\subsubsection{Domains set $E$}
Flexible resource $V$ can be set to either controllable or uncontrollable.
Therefore, the flexibility request can be decomposed into multiple domains.
These settings are decided by the individual prosumer and can vary from time to time depending on occupants' preferences and behavioural patterns.
Further, the controllable resources can be decomposed into subdomains.
These domains and their subdomains are illustrated as follows:
\begin{equation}
    \begin{aligned}
        df_{t}^{s,q,d}(CR)=\sum_{i=1}^m (df_{t}^{s,q,d}(DG)+df_{t}^{s,q,d}(HW)+\\
        df_{t}^{s,q,d}(HVAC)+df_{t}^{s,q,d}(ESS)) 
    \end{aligned}
\end{equation}
Where $m$ is the total number of controllable resources $CR \subseteq V$, and $df_{t}^{s,q,d}$ is the total DF cost of using flexible generation $DG$ systems such as rooftop solar panels and wind turbines, hot water $HW$ systems such as thermal solar and heat pumps, heating, ventilation, air conditioning and cooling $HVAC$ systems, and Energy storage systems $ESS$ such as batteries, Electric Vehicles (EV), and thermal storage respectively. 

\subsubsection{Constraint set $C$}
The objective function is subject to a set of constraints $C$.
Each constraint $c \in C$ is a high-order relationship $(s,\rho)$, such that the scope $s=\{v_{1},v_{2},v_{3}\}$ is a set of flexible resources, and $\rho$ is the assignment function that can vary the setpoints of all flexible resource in the scope $s: \forall v \in V$.
For instance, \Cref{eq:control_islanding} illustrate an objective function to provide flexibility service for \textbf{Control Islanding} \cite{Chuang2008}.
\begin{equation}
    \begin{aligned}
        min\int_{DF_{t}^{s,q,d}}^{\$} = \sum_{t\in T} (\sum_{g\in DG}g_{t}^{out}+\sum_{h\in HW}h_{t}^{OFF} \\
        +\sum_{\nu\in HVAC}\nu_{t}^{OFF}+\sum_{\mu\in ESS}\mu_{t}^{-ve})
    \end{aligned}
    \label{eq:control_islanding}
\end{equation}
This objective function requires the prosumers to decrease their energy consumption and increase their local production by varying their flexible resources based on the following constraint functions:
\begin{itemize}
    \item \textbf{$g_{t}^{out}$:} increases the generation output of flexible generation unit $g \in DG$ during period $t$.
    \item \textbf{$h_{t}^{OFF}$:} switch off the water heater system $h \in WH$ during period $t$.
    \item \textbf{$\nu_{t}^{OFF}$:} switch off the HVAC system $\nu \in WH$ during period $t$.
    \item \textbf{$\mu_{t}^{-ve}$:} discharge the energy storage system $\mu \in ESS$ during period $t$.
\end{itemize}
Once the flexibility period $T$ elapses, all flexible resources return to the baseline operation. 
A baseline is the operation mode expected by the flexible resource if a constraint had not occurred in response to a flexibility request.
Moreover, all interaction events generated by the dFASC are recorded on the blockchain ledger for transparency and auditability.

\section{Evaluations and Discussions} \label{sec:security_evaluation}
In this section, we present a discussion on the security and performance evaluations of PlexiChain.
To evaluate the potential of cyberattacks on the PlexiChain framework, we adopted the widely used STRIDE threat model \cite{Hussain2014}. 
Then, using the Hyperledger Caliper\footnote{\url{https://hyperledger.github.io/caliper/}} framework, we benchmark the performance of the certificate-based PKI in HARB against the NFT-based PKI in PlexiChain. 

\subsection{Security Evaluations}
The security evaluation process comprises of three main steps, namely system modelling,  threats identification, and threat analysis.
We detail each step as follows:



\subsubsection{System Modelling} \label{sec:system_model}
\begin{figure}[!h]
    \centering
    \includegraphics[width=\linewidth]{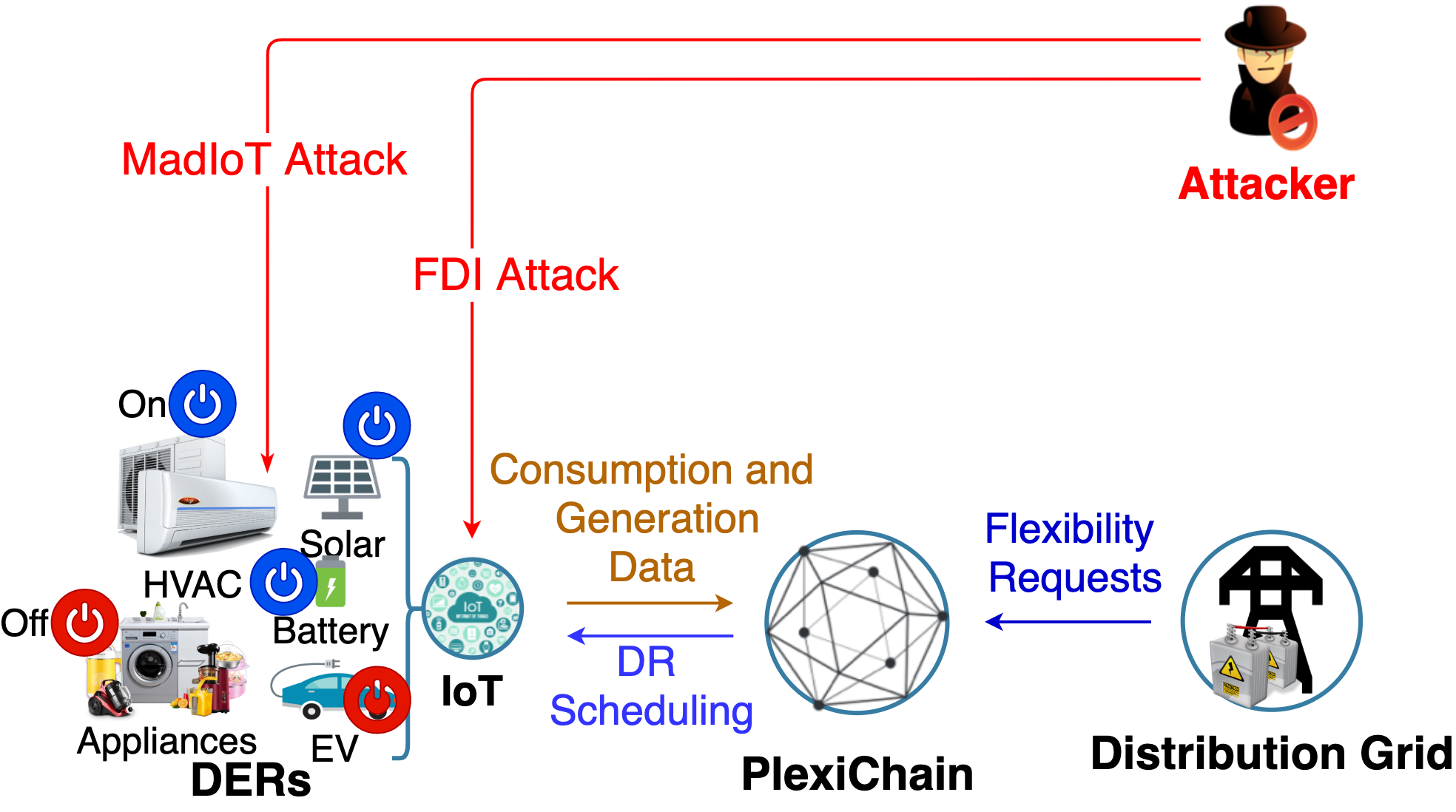}
    \caption{Decentralised Aggregator Model}
    \label{fig:threat_model}
\end{figure}

Essential, modelling a system under analysis is used to identify the system components and assets that need to be secured \cite{Hussain2014}.
\Cref{fig:threat_model} depict the interactions between components in the decentralised aggregator model.
In the proposed decentralised aggregator model, the flexible resources and the distribution grid interact directly through the PlexiChain framework without the need for a TTP.
The PlexiChain framework uses the dFASC chaincode to process the flexibility requests from the distribution grid.
Similarly, the PlexicHain framework uses the dFASC chaincode to process the flexibility bids from the flexible resources.

\subsubsection{Evaluation of security vulnerabilities, threats and attacks}
A vulnerability is a weakness in the system caused by poor design, configurations mistake, or insecure coding practices.
A threat is a possible harm that may occur if an internal or external attacker exploits vulnerabilities in the design of a system.
Consequently, an attack is a potential for loss, damage or destruction of an asset when an attacker exploits a vulnerability or enacts a threat  \cite{Hussain2014}.
We adopted the STRIDE threat model \cite{Hussain2014} to assess and evaluate the system model illustrated in \Cref{fig:threat_model} for vulnerabilities, threats, and attacks.
\Cref{tab:STRIDE_threat_model} presents a summary of each threat and the security property violated.
Then, using an impact metric of complete, partial and none, we analysed the possible threat level impacted by an attacker.
\textbf{Complete} indicates a total loss of the security property violated.
\textbf{Partial} indicates a slight loss on the security property violated.
\textbf{None} indicates no impact on the security property violated.
Evaluation results summarised in \Cref{tab:vulnerabilities}.
\begin{itemize}
    \item \textbf{IoT Network:} 
    Most IoT devices used in the built environment lack the appropriate safeguards (e.g. antivirus and anti-spyware software, firewalls, and intrusion detection systems) or have back-doors and hard-coded passwords placed on them by manufacturers \cite{Soltan2018}.
    The lack of safeguards on IoT devices exposes the system to threats to confidentiality, integrity and availability. 
    
    \item \textbf{Flexible resources:}
    The PlexiChain to flexible resource control and communication protocol implements the OpenADR communication standard \cite{DOE2021}.
    However, OpenADR standards implemented over the older version of TLS (e.g. v1.3) are vulnerable to spoofing, tampering and information disclosure security threats \cite{Hussain2014}.
    
    \item \textbf{Flexibility market:} 
    Recall that flexibility settlements are carried outside the PlexiChain system in an incentive-based market model. 
    This approach is vulnerable to repudiation attacks as it assumes that each participant can be trusted.
\end{itemize}
\begin{table*}[!ht]
    \centering
    \caption{STRIDE Threats Categories}
    \label{tab:STRIDE_threat_model}
    \resizebox{\linewidth}{!}{%
    \begin{tabular}{llllll}
    \toprule
    \multirow{2}{*}{\textbf{}} &    \multirow{2}{*}{\textbf{Threats}} & \multirow{2}{*}{\textbf{Property Violated}} &
    \multirow{2}{*}{\textbf{Threat Definition}}
    \\ {}&{}&{} \\
    \midrule
    \textbf{S}  & Spoofing & Authentication & Impersonating something or someone else  \\
    \textbf{T} & Tampering & Integrity & maliciously modifying data at collection, transmission, or storage  \\
    \textbf{R} & Repudiation & Non-repudiation & Claiming that you did not perform a certain action \\
    \textbf{I} & Information Disclosure  & Confidentiality & Exposing information to someone who is not authorised to access it \\
    \textbf{D} & Distributed Denial of Service (DDoS)  & Availability & Deny or degrade resource needed to provide service \\
    \textbf{E} & Elevation of Privilege (EoP) & Authorisation & Allowing someone to do something without proper authorisation \\
     \bottomrule
    \end{tabular}}
\end{table*}
\begin{table*}[!ht]
    \centering
    \caption{A summary of DR System components and subsystems vulnerabilities, potential threats and their impact levels.}
    \label{tab:vulnerabilities}
    \resizebox{\linewidth}{!}{%
    \begin{tabular}{llllll}
    \toprule
    \multirow{2}{*}{\textbf{Components and Sub-systems}} & \multirow{2}{*}{\textbf{Vulnerabilities}} &
    \multirow{2}{*}{\textbf{Threats}} &\multicolumn{3}{c}{\textbf{CIA Security Impact Metric}} \\
    {}&{}&{}&
    \multicolumn{1}{c}{Confidentiality} & \multicolumn{1}{c}{Integrity}     & \multicolumn{1}{c}{Availability}     \\ \midrule
    IoT Network & Backdoors  \& Hard-coded passwords  & Spoofing, Tampering \& DDoS & Complete & Complete & Complete \\
    OpenADR Standard & Outdated TLS schemes  & Spoofing, EoP \& Tampering & Partial & Complete & Complete \\
    Digital marketplaces & Flawed contractual agreements & Repudiation \& Info. disclosure & None & Complete & None  \\
     \bottomrule
    \end{tabular}}
\end{table*}

\subsubsection{System Analysis}
Lastly, we performed a quantitative risk analysis of the proposed system model based on the most common attacks in flexibility markets.
These attacks include: 

\begin{itemize}
    \item \textbf{False Data Injection (FDI):}
    FDI attacks are common measurement integrity attacks where an adversary modifies the stored measured states or directly manipulates the measured states in the IoT devices \cite{Qi2017a}.
    As illustrated in \Cref{fig:threat_model}, an attacker can perform an FDI by exploiting tampering threats and IoT vulnerabilities on the network layer.

    \item \textbf{Manipulation of Demand via IoT (MadIoT):}
    MadIoT attacks are common integrity attacks on control commands, and signals \cite{Soltan2018}.
    As illustrated in \Cref{fig:threat_model}, the attacker can synchronously switch on/off compromised high-wattage devices simultaneously, causing a sudden increase or decrease in electricity demand.
\end{itemize}
For our experiments, we used a real-world consumption and generation dataset collected from an office building in Newcastle, Australia, for one year with a sampling period of 30 minutes.
The dataset characteristics are summarised below.

\begin{figure*}[!th]
    \centering
    \subfigure[Normal HVAC energy consumption]{
        \includegraphics[width=.3\linewidth]{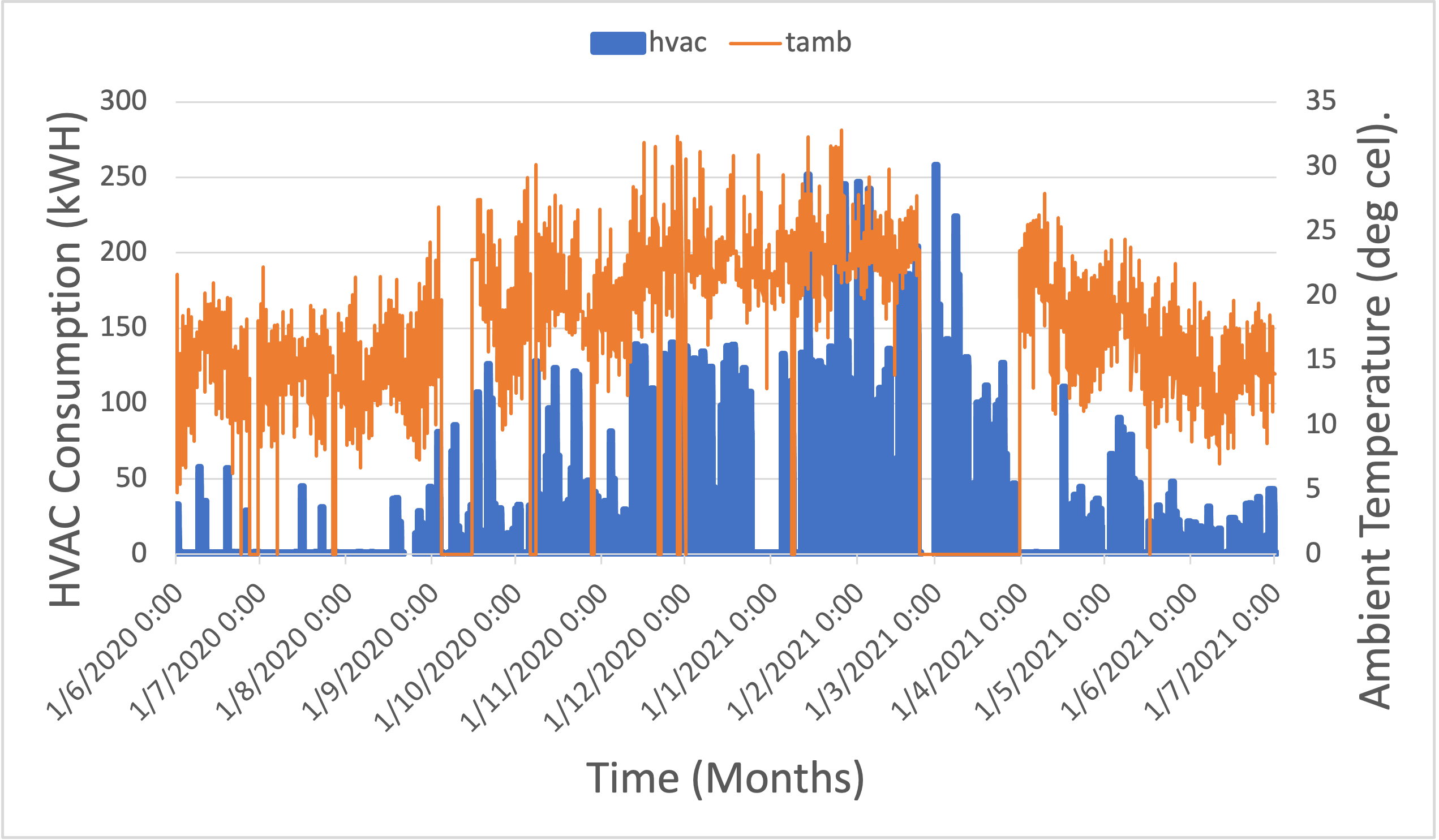}
        \label{fig:evaluation_actual}
    }
    \subfigure[HVAC energy consumption after MadIoT attack]{
        \includegraphics[width=.3\linewidth]{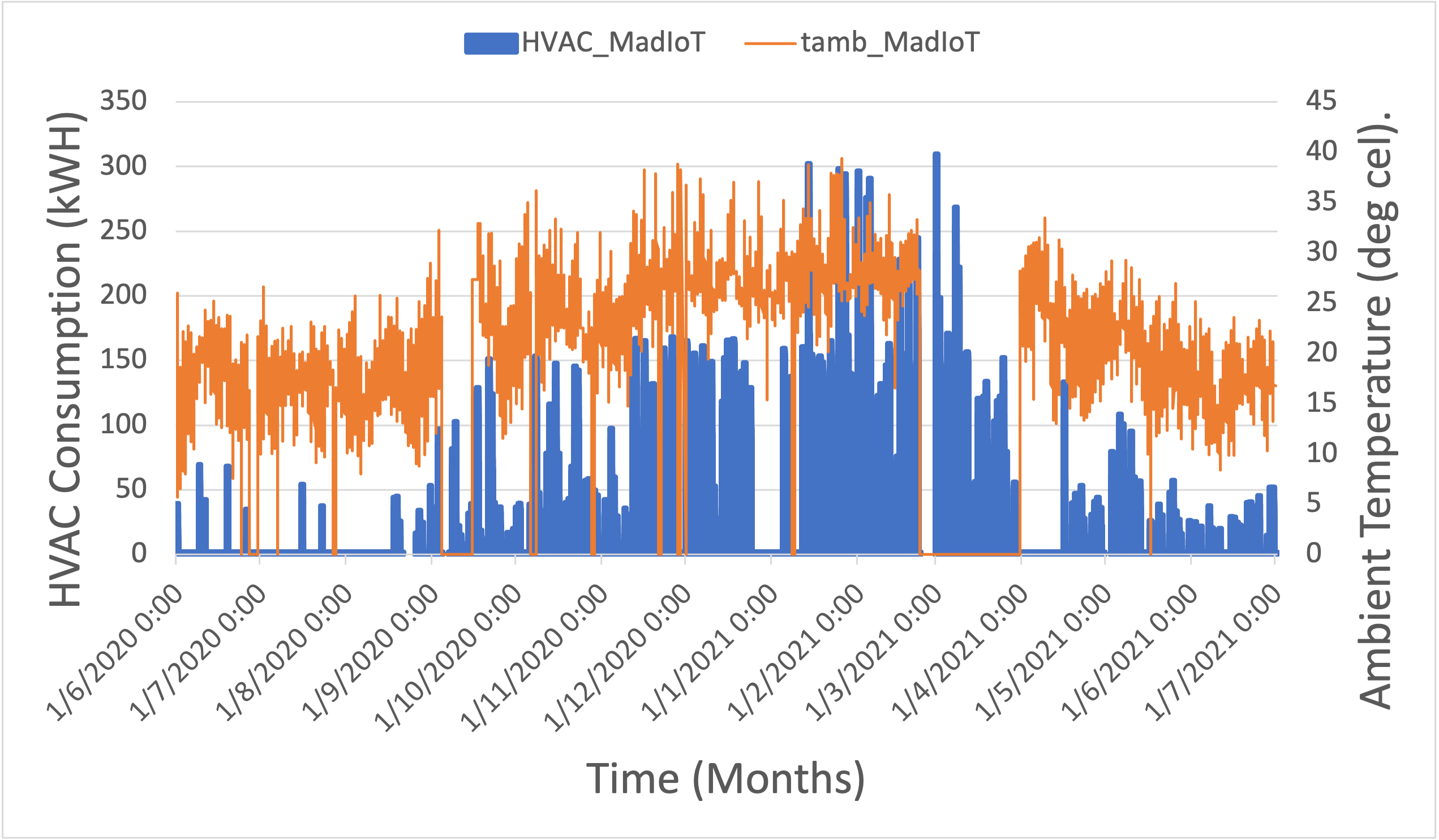}
        \label{fig:evaluation_madiot}
    }
    \subfigure[Building's net power after FDI attack]{
        \includegraphics[width=.3\linewidth]{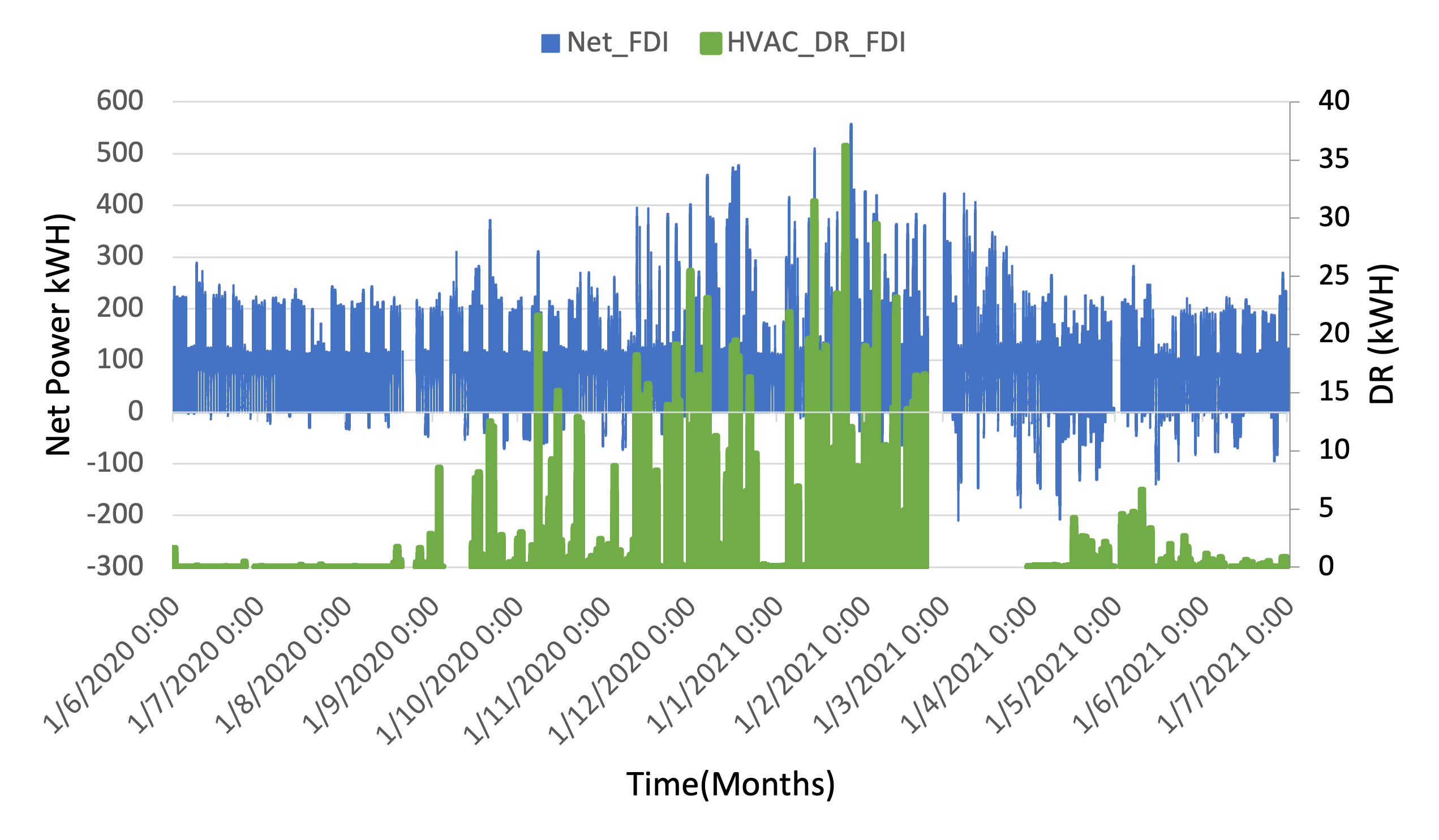}
        \label{fig:evaluation_fdi}
    }
    \caption{MadIoT attack evaluation}
    \label{fig:MadIoT_evaluations}
    
\end{figure*}

\begin{itemize}
    \item \textbf{Time:} Australian Eastern Daylight Time $t$ at 30 minutes intervals.
    \item \textbf{Net:}  site net power in kW. Net power is a computed state given by consumption minus generation measured states.
    Net power is negative when the site generates more power than it consumes.
    \item \textbf{tamb:} ambient temperature as measured by a sensor on-site (degrees Celsius).
    \item \textbf{HVAC:}  measured consumption power (kW) of HVAC systems in the Building used for heating, cooling, air-conditioning etc. 
    This excludes AHU (air handling unit) power.
    \item \textbf{hvac\_demand\_res:}  is a theoretical estimate of the Building's DR available. 
    This has been computed using the methodology described in this literature \cite{Goldsworthy2020}.
\end{itemize}
We used the dataset to demonstrate the aforementioned FDI and MadIoT attacks.
In this case, a FDI attacker can only manipulate the data during collection and transmission.
\Cref{eq:FDI_attack_profile} shows the mathematical representation for the attack profile generation \cite{Madabhushi2020}.
\begin{equation} \label{eq:FDI_attack_profile}
    \begin{bmatrix} g_{1}\\ g_{2}\\ .\\ .\\ g_{t} \end{bmatrix} + \begin{bmatrix} d_{1}\\ d_{2}\\ .\\ .\\ d_{t} \end{bmatrix} = \begin{bmatrix} c_{1}\\ c_{2}\\ .\\ .\\ c_{t} \end{bmatrix}
\end{equation}
Where $g_{t}$ is the original measured states at time $t$, $d_{t}$ is the added value at time $t$, and $c_{t}$ is the resultant attack profile at time $t$ obtained by adding the attack value to the measured states.
We apply the formula presented in \Cref{eq:FDI_attack_profile} on the dataset's Net power column with $d_{t}=2\%$. 
The evaluation results presented in \Cref{fig:MadIoT_evaluations} show that increasing the net power output increases the estimated demand flexibility output.

To generate a MadIoT attack profile, we used \Cref{eqn:gain}, a mathematical formulation for generating MadIoT attacks as proposed in \cite{Madabhushi2020}.  
\begin{equation}
    G_{t} = \sum_{i=1}^{t}d_{i}
    \label{eqn:gain}
\end{equation}
Where $G_{t}$ is the HVAC gain in kilowatts at time $t$, $i$ ranges from the first minute of attack until the end of attack time $t$ in minutes and $d_{i}$ is sensor reading at time $0<i<t$.
We considered a scenario where an attacker increases the sensor reading by $2\%$.
The MadIoT evaluation result illustrated in \Cref{fig:MadIoT_evaluations} shows that increasing the ambient temperature would lead to a reduced demand flexibility estimation. 

The proposed PlexiChain framework addresses the STRIDE security threats outlined in \cref{sec:security_evaluation}, as follows:

\begin{itemize}
    \item \textbf{Spoofing:}
    To protect against spoofing attacks, our framework requires all entities to have public-private key pairs for authentication.
    The public-private key pair is bound to an NFT token to verify each entity's identity on the network.
    \item \textbf{Tampering:}
    The proposed PlexChain framework utilises blockchain and digital signatures to sign and encrypt data from IoT devices for data integrity verification.
    Each time data is retrieved from the blockchain ledger, the signature is used to validate data integrity.
    \item \textbf{Non-Repudiation:}
    To provide non-repudiation, the PlexiChain framework implements the business logic through smart contracts such as the dFASC, dTAM and the NFT SC discussed in \Cref{sec:plexichain}, without the need for trust and TTP. 
    Additionally, all actions performed by each entity are immutably recorded on the blockchain ledger.
    \item \textbf{Information disclosure:}
    The proposed NFT-based PKI eliminates the need for TTP such as the CA, which is susceptible to private information leakage.
    The PlexiChain framework uses the NTF smart contract for identity verification and does not need to share private information with a TTP such as the CA.

    \item \textbf{DDoS:}
    Plexichain ensures \textit{availability} through distributed ledger storage. 
    To compromise a blockchain network, one would require to compromise 51\% of the available node.

    \item \textbf{Elevation of Privilege:}
    The NFT token implements an access control property with a list of access rights that can be granted to a certain entity on the PLexiChain system.
    This information cannot be changed without properly issuing a new NFT token.
\end{itemize}

\begin{figure*}[!ht]
    \centering
    \subfigure[Memory Footprint]{
        \includegraphics[width=.45\linewidth]{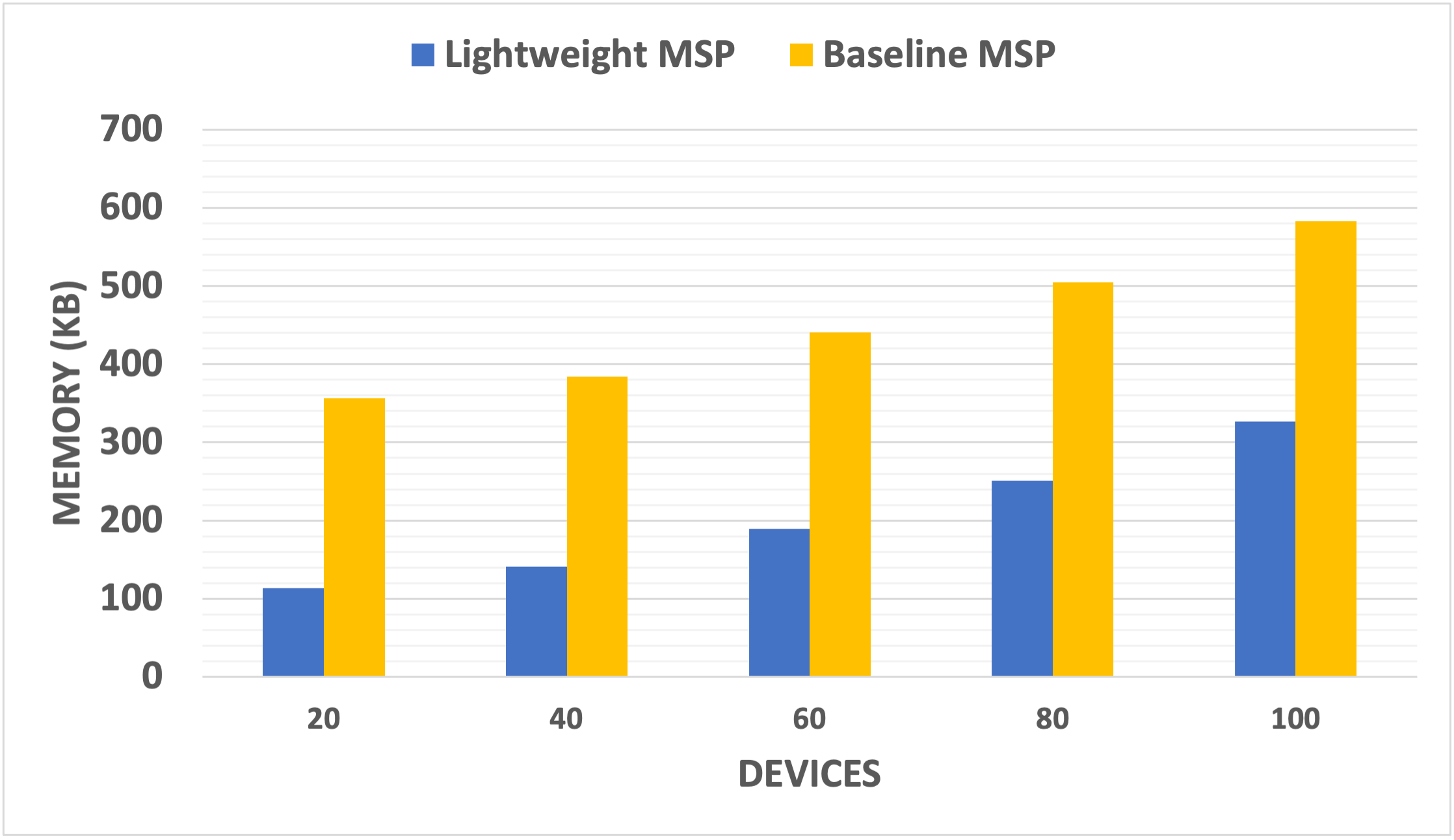}
        \label{fig:eval_memory}
    }
    \subfigure[Throughput and Latency]{
        \includegraphics[width=.45\linewidth]{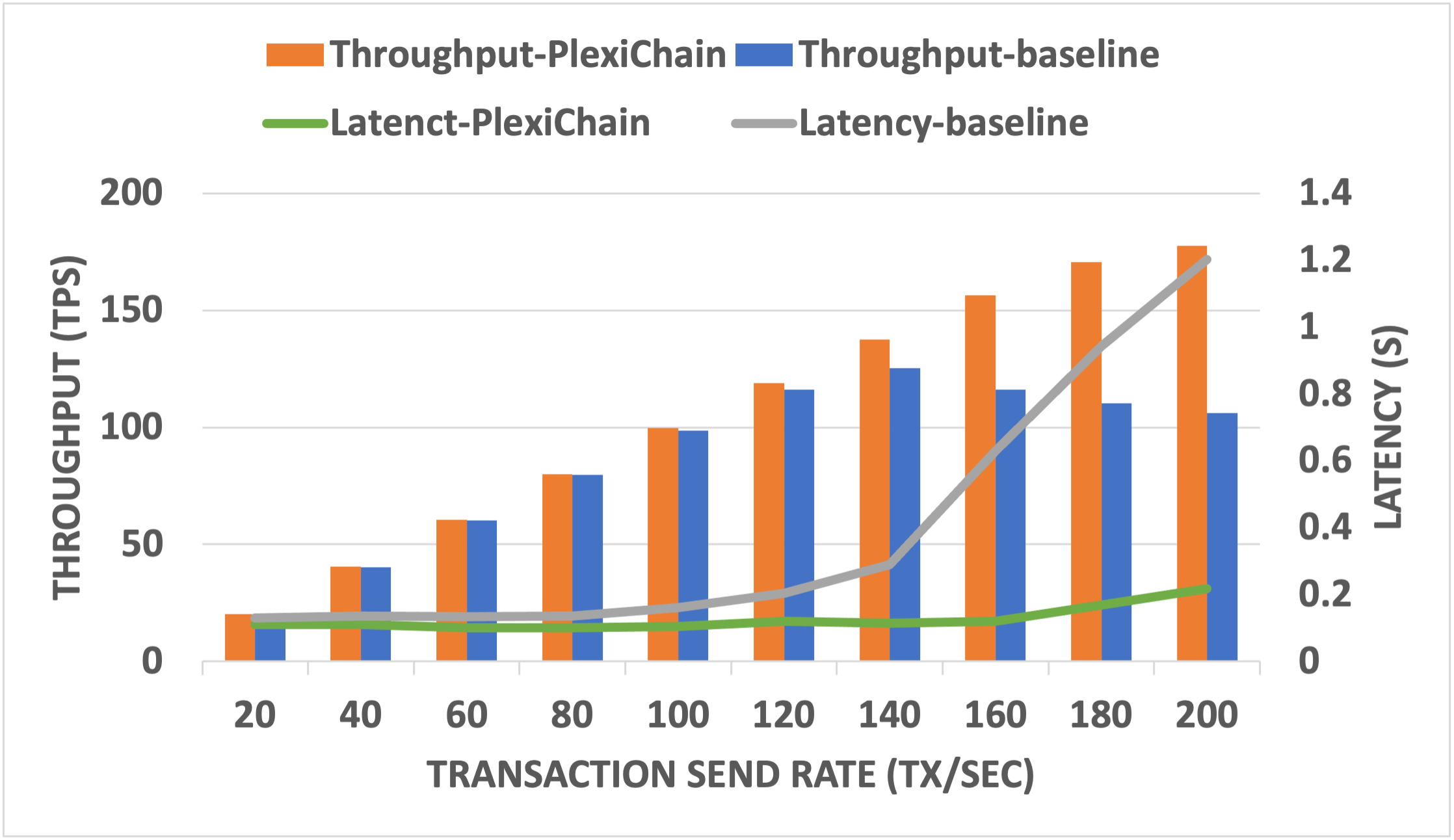}
        \label{fig:eval_throughput_latency}
    }
    \caption{Performance Evaluation}
    \label{fig:results}
\end{figure*}

\subsection{Performance Evaluation} \label{sec:performance_evaluations}
This section covers the performance evaluation of PlexiChain.
First, we describe the experimental setup and the underlying technologies used to implement the proposed solutions.
Then, using the Hyperledger Caliper\footnote{\url{https://hyperledger.github.io/caliper/}} framework, we evaluated the computation and communication cost of blockchain using the NFT-base PKI component and benchmarked it against the certificate-based PKI.


\subsubsection{Experiment Setup} \label{sec:implementation}
We have deployed: (i) 6 Raspberry PI 4's with 8GB LPDDR4-3200 SDRAM, a Broadcom BCM2711, and Quad-core Cortex-A72 (ARM v8) 64-bit system on chip with 1.5 GHz frequency.
(ii) A MacBook Pro with 8 GB memory and Dual-Core Intel Core i5 of 2.3 GHz frequency. and 
(iii) A Digital Ocean droplet with 8 GB memory and 4vCPUs of 2.4 GHz frequency each.
Each Raspberry PI represents a smart meter equipped with a local controller to relay energy consumption data from flexible resources and DF scheduling commands from the PlexiChain platform.
The MacBook Pro and  Digital Ocean droplet represent high-performing nodes owned by the flexibility customers (DSO/TSO) for transacting DF in flexibility markets.
These devices form the blockchain nodes on which the PlexiChain network is deployed in a Docker virtual environment.

Docker\footnote{\url{https://docs.docker.com/get-started/overview/}} is a set of platform-as-a-service products that use OS-level virtualisation to deliver microservices in packages called containers.
First, we describe and configure each PLexiChain component (i.e. peers, orderers, notaries and committers) as a micro-service in a docker-compose (version 3.0) script.
Compose\footnote{\url{https://docs.docker.com/compose/}} is a tool for defining and running multiple docker containers.
At the application layer, we used Node-Red\footnote{\url{https://nodered.org/}} to implement the DF Trading DApp.
Node-Red is a programming tool for connecting hardware devices (e.g. sensors, smart meters, and flexible resources), APIs and other online services.
Lastly, all smart contracts were implemented using the golang programming language and deployed on the endorsing peers.

\subsubsection{Results and Discussions}
This section evaluates the PlexiChain framework and benchmarks it against the certificate-based HARB framework as our baseline.
Recall that one of the key contributions of PlexiChain is to reduce the high computation and communication costs incurred during certificate management in PKI.
To evaluate this, we compare our framework's memory footprint, transactions throughput, and computing latency with that of the baseline.
Throughput is the rate at which transactions are processed after being received, while latency is the time taken from when a transaction is issued to the time it is committed \cite{Karumba2020}.
We use Hyperledger Caliper, a benchmarking tool for blockchain frameworks, to evaluate the two Systems Under Test (SUT). 
The results are presented as follows:

\noindent \textbf{Memory footprint:}
    To evaluate the memory footprint, we implement the key generation phase algorithms of PlexiChain and the baseline in the Caliper's workload configuration file.
    The workload configuration file implements the business logic to emulate a SUT client.
    We then configure the caliper benchmark configuration file to send a workload at the rate of 20, 40, 60, 80 and 100 to the SUT.
    The results show that PlexiChain reduces the amount of memory used by about 33.33\%, as shown in \Cref{fig:eval_memory}.
    The NFT-based PKI only stores the partial private keys for each device required to generate the private-public key pair, while the baseline PKI stores both the certificate and the public-private key pair for all devices.
    
\noindent \textbf{Throughput and Latency:}
    We implement the transaction signing phase of both PlexiChain and baseline to evaluate throughput and latency in a new caliper workload configuration file.
    Then, we configure the benchmark configuration file to execute the workload by sending transactions to the SUT at a send rate of 20, 40, 60, and thorough 200 transactions per sec (tps). 
    \Cref{fig:eval_throughput_latency} shows a comparison of latency and throughput between PlexiChain and the baseline. 
    Throughput results show that for the given transactions range, the baseline reaches saturation point at 120tps,  while that of PlexiChain is higher than 170tps.
    PlexiChain improves throughput because the transaction size is much smaller than the baseline, including a certificate for signing key verification.
    Similarly, latency in PlexiChain significantly reduces compared to the baseline, especially at higher transaction rates.
    Note that latency increases exponentially after the baseline reaches the throughput saturation point.
    This is because Caliper does not have a queuing mechanism, and thus the transactions start to fail.

\section{Conclusions} \label{sec:conclusions}
In this work, we have proposed PlexiChain: a secure blockchain-based flexibility aggregator framework for the decentralised aggregator model.
Our framework utilises the blockchain's salient features (i.e. decentralisation, immutability, and security) to address the outlined security and trust challenges.
Using a STRIDE threat model, we identified security threats and vulnerabilities that can be exploited to compromise the confidentiality, integrity and availability of DR systems.
Additionally, we argued that using a CL-PKC base on NFTs would help reduce the computation and communication overhead arising from certificate management in the certificate-based PKC.
Simulation results show that the proposed PlexiChain framework reduces the computation and communication cost by increasing the network throughput and reducing the overall network latency.

\bibliographystyle{unsrt}

\bibliography{references}

\end{document}